\UseRawInputEncoding
\documentclass[reprint,amsmath,amssymb,aps,nofootinbib,superscriptaddress]{revtex4-2}
\usepackage{graphicx}
\usepackage{dcolumn}
\usepackage{bm}
\usepackage{color}
\usepackage{empheq}
\usepackage[colorlinks]{hyperref}
\usepackage[dvipsnames]{xcolor}
\usepackage{orcidlink}
\usepackage{tikz}
\usepackage{tikz-feynman}
\usetikzlibrary{patterns}
\usetikzlibrary{decorations.markings}
\usetikzlibrary{arrows.meta}
\tikzset{
    mid arrow/.style={
      postaction={
        decorate,
        decoration={
          markings,
          mark=at position 0.6 with {\arrow{Latex[scale=1]}}
        }
      }
    }
  }
\definecolor{hgreen}{rgb}{0.0, 0.44, 0.0}
\definecolor{carnelian}{rgb}{0.65, 0.11, 0.11}
\definecolor{shadecolor}{rgb}{0.92,0.92,0.92}
\hypersetup{colorlinks=true, linkcolor=carnelian, filecolor=carnelian, urlcolor=carnelian, citecolor=carnelian, linktoc=page}

\def\be{\begin{equation}}
\def\ee{\end{equation}}
\def\d{\partial}
\def\locfnl{f_\text{NL}^\text{loc}}

\def\S{\mathcal{S}}

\def\k{\textbf{k}}
\def\q{\textbf{q}}
\def\p{\textbf{p}}
\def\x{\textbf{x}}
\def\y{\textbf{y}}
\def\n{\hat{\mathbf{n}}}

\begin{document}

\title{Imprints of Large-Scale Structures in the Anisotropies of the Cosmological Gravitational Wave Background}

\author{Rafael Bravo\,\orcidlink{0000-0003-1598-9061}}
\email{rafael.bravo@uc.cl}
\affiliation{Instituto de F\'{i}sica, Pontificia Universidad Cat\'{o}lica de Chile, Avenida Vicu\~na Mackenna 4860, Santiago, Chile}

\author{Walter Riquelme\,\orcidlink{0000-0002-5341-6793}}
\email{walter.riquelme@ictp-saifr.org}
\affiliation{ICTP South American Institute for Fundamental Research, IFT-UNESP, S\~{a}o Paulo, SP 01440-070, Brazil}
\thanks{Authors are listed alphabetically.}

\begin{abstract}
We compute the cross-correlation between the anisotropies of the cosmological gravitational wave background (CGWB) and the galaxy density contrast. 
We show that the cross-correlation is non-zero due to the {\it late} integrated Sachs-Wolfe (ISW) effect experienced by tensor modes.
We study the detection prospects of the cross-correlation signal against cosmic variance (CV), and in the light of incoming LSS and GW surveys, where we found that the signal under certain conditions could be distinguishable from noise.
In addition, by considering a CGWB sourced by scalar-induced gravitational waves, and the inclusion of a scale-dependent galaxy bias, we use the cross-correlation to forecast local primordial non-Gaussianity, where we find $\sigma(\locfnl)\sim10$ for CV only and a LSST-like survey.
Moreover, by combining the Fisher information of CGWB$\times$LSS with LSS, we are able to improve the constraints by 4\% compared to an LSS-only analysis. 
Our results imply that a cross-correlation between GW anisotropies and LSS can indeed come from a stochastic background of cosmological origin and could be used to distinguish it from an astrophysical one.
\end{abstract}
\maketitle

\section{Introduction}
The discovery of the fossil photon bath permeating the entire universe, and the subsequent detection of its anisotropies, earned cosmology the title of precision science.
Today, we may be on the verge of a similar breakthrough.
The direct detection of gravitational waves (GW) a decade ago \cite{LIGOScientific:2016aoc} established a new paradigm to observe the universe.
Such a milestone provides a solid foundation for the expectation that the universe is also permeated with a stochastic gravitational wave background (SGWB) \cite{Maggiore:1999vm, Caprini:2018mtu}, as several pulsar timing array (PTA) collaboration reports seem to indicate \cite{NANOGrav:2023gor, Xu:2023wog, EPTA:2023fyk, Reardon:2023gzh}.
The magnitude of the SGWB is quantified in terms of its energy density spectrum $\Omega_\text{GW}(q)$, where, although there are bounds on its amplitude \cite{KAGRA:2021kbb}, a plethora of astrophysical and cosmological sources can contribute to it, each of them with its own spectral shape \cite{Figueroa:2023zhu}. 
Therefore, an important open problem is how to distinguish between astrophysical and cosmological sources contributing to the SGWB \cite{Ellis:2023oxs, NANOGrav:2023hvm}. 
For such a purpose, one can consider the fact that the SGWB would not look the same in every direction in the sky, {\it i.e.,} its energy spectrum would display anisotropies, $\Omega_\text{GW}(q,\n)$ \cite{Contaldi:2016koz}. 
Hence, different sources would exhibit distinctive signatures of their anisotropies \cite{Li:2023qua, LISACosmologyWorkingGroup:2022kbp}. 

Beyond including anisotropies, in recent years, it has been proposed that cross-correlations of the SGWB with the cosmic microwave background (CMB) and the large-scale structure (LSS) of the universe would provide further hints about its origin \cite{Ricciardone:2021kel, Braglia:2021fxn, Perna:2023dgg, Zhao:2024gan, Scelfo:2018sny, Bosi:2023amu, Semenzato:2024mtn, Pedrotti:2025tfg, Cusin:2025xle}.
In particular, in \cite{Semenzato:2024mtn} it has been claimed that the non-zero correlation of the SGWB with LSS would support astrophysical sources as the primary component of the SGWB, and therefore, allowing the SGWB to be exploited as an LSS tracer \cite{Libanore:2020fim}.
In this article, we argue that this claim must be taken with caution.

Even though most of the anisotropies of the cosmological gravitational wave background (CGWB) are generated at very early times, there is a contribution that is gravitationally induced at late times. 
When the growth of structure slows down, as happens in a dark energy-dominated ($\Lambda$D) universe, the free streaming of GWs passing through decaying gravitational potentials can induce a net energy gain on them, leading to anisotropies at low redshifts.
This is the so-called {\it late} integrated Sachs-Wolfe (ISW) effect \cite{Sachs:1967er}.
Since the same gravitational potentials also serve as seeds for the LSS of the universe, it is then expected that the cross-correlation between CGWB anisotropies and an LSS tracer to be non-zero due to this effect, as happens with CMB$\times$LSS correlations \cite{Crittenden:1995ak, Giannantonio:2008zi, Cooray:2001ab, Ho:2008bz, Giannantonio:2013uqa}. 

The main purpose of this article is to show that the CGWB$\times$LSS correlation indeed is non-zero and, as expected, dominated by the ISW effect. 
We show this, in close analogy with the CMB, {\it i.e.} by computing the correlation between the anisotropies of the CGWB with the galaxy-density contrast serving as LSS tracer of the local gravitational potential. 
On the one hand, in order to provide a background of cosmological origin, we consider GWs sourced by large primordial fluctuations that re-enter the horizon during the radiation-dominated (RD) universe, the so-called scalar-induced gravitational waves (SIGWs) (see \cite{Domenech:2021ztg} for a comprehensive review).
It is known that, aside from sourcing GWs, large primordial fluctuations may collapse to form primordial black holes (PBHs) \cite{Sasaki:2018dmp, Carr:2023tpt} as well as exhibit non-negligible non-Gaussianity \cite{Cai:2018dig}. 
Moreover, if these non-Gaussianities are of the local type, they enable a coupling between short- and long-wavelength fluctuations, allowing the latter to redistribute the power spectrum of the former across large scales, so that they could emerge as anisotropies in the gravitational wave background \cite{Wang:2023ost}.
Non-Gaussianities of the local type are usually parametrized as 
\be \label{local-ansatz}
\zeta (\x) = \zeta_g(\x) + \locfnl\left(\zeta_g^2(\x)-\langle\zeta_g^2\rangle\right),
\ee
where $\zeta$ is the non-Gaussian primordial curvature perturbation arising from quantum fluctuations during inflation, expanded around a Gaussian perturbation $\zeta_g$, and the constant parameter $\locfnl$ quantify the deviations from Gaussian statistics. 
It is worth mentioning that any detection of $\locfnl \neq 0$ would rule out single-field inflation models with {\it adiabatic} long-wavelength fluctuations, such as slow-roll (SR) and ultra-slow-roll (USR) \cite{Tanaka:2011aj, Pajer:2013ana, Pimentel:2013gza, Bravo:2020hde}. Hence, its detection is of utmost importance, since it can provide important insights about the dynamics of inflation.

On the other hand, from the LSS standpoint, $\locfnl$ induces the same coupling of long and short modes which influences the structure formation by introducing a scale-dependent effect on the galaxy bias. This effect can be used to constrain $\locfnl$ \cite{Chaussidon:2024qni, Riquelme:2022amo}. However, since the constraining power of these surveys is insufficient to claim a detection, an active effort has started to improve the constraints towards the threshold of detection. 
A promising strategy to improve the constraints is to combine multiple tracers of the underlying matter distribution \cite{Giannantonio:2013uqa, Abramo:2013awa}.

By combining non-Gaussian SIGWs and the effect of the scale-dependent bias on the galaxy-density contrast, we compute the GW$\times$LSS correlation and study its detection prospects for future GW and LSS surveys. We then take advantage of the possibility of detection to analyze whether it can be used as a window to constrain $\locfnl$. 

This article is organized as follows: In Sec. \ref{sec-II} by following a Boltzmann approach, we compute the CGWB anisotropies produced by SIGWs. 
In Sec. \ref{sec-III} we introduce the basics of the LSS tracers that we will later use in Sec. \ref{sec-IV} to compute the CGWB$\times$LSS cross-correlation, where we show that the dominant contribution comes from the ISW effect, and that it is sensitive to $\locfnl$. 
In Sec. \ref{sec-V} we discuss the detection prospects of this signal by performing an SNR analysis. Once their detectability is assessed, we perform a Fisher forecast for $\locfnl$ using the cross-correlation combined with the autocorrelation of galaxies.
Finally in Sec. \ref{sec-VI} we offer some concluding remarks.
\section{CGWB Anisotropies}\label{sec-II}
Gravitational waves are produced by astrophysical or cosmological sources and then propagate through an expanding and inhomogeneous universe until they reach a detector. Both sources, and their subsequent interaction with cosmic structures, can introduce directional variations in the energy flux they carry. These variations later unfold as anisotropies in the observed GW energy density spectrum \cite{Contaldi:2016koz}.

To quantify the anisotropies, one uses a statistical approach. The population of gravitational wave modes, each identified by a spacetime position $x^\mu$ and momentum $p^\mu$, is described through a distribution function $f(x^\mu, p^\mu)$, from where one can compute the GW energy density $\rho_\text{GW}$.  
Effective theory considerations \cite{Isaacson:1968hbi} allows us to consider gravitons as massless and collisionless modes propagating along null geodesics of the perturbed spacetime 
\be \label{pert-FLRW}
ds^2  = a^2(\eta)\left[-(1+2\Phi)d\eta^2 + \left(1-2\Phi\right)\delta_{ij}dx^i dx^j\right],
\ee
where $a(\eta)$ is the FLRW scale factor as a function of the conformal time $\eta$, $\Phi(\eta, x^i)$ is a scalar linear perturbation 
and we have neglected linear vector and tensor perturbations.
Since we are interested in GWs of cosmological origin, the Boltzmann equation for $f(\eta,\x,\q)$, with the above mentioned considerations, up to linear order in $\Phi$ reads,
\be \label{pert-boltzmann}
\frac{\d f}{\d \eta} + n^i \frac{\d f}{\d x^i}+\left[\frac{\d \Phi}{\d \eta} - n^i \frac{\d\Phi}{\d x^i}\right] q \frac{\d f}{\d q} = 0, 
\ee
where $\n \equiv \hat{\p}$ is the unit vector along the graviton propagation direction, and $q\equiv a(\eta) |\mathbf{p}|$ its comoving momentum. By expanding the distribution function up to linear order in perturbations, $f(\eta, \x, \q) = \bar f(\eta, q) - \d\bar{f}/\d\ln q \,\Gamma(\eta, \x, \q)$, one can solve the Boltzmann equation at zeroth and linear order respectively. 
The zeroth order equation, $\d \bar f / \d \eta = 0$ is solved by a distribution that is function only of the comoving momentum $\bar{f}(q)$,  indicating that the homogeneous isotropic background does not evolve in time.
While the first order Boltzmann equation in Fourier space,
\be \label{Boltzmann-1}
\frac{\d \Gamma}{\d \eta} + i k \mu \Gamma = \frac{\d \Phi}{\d \eta} - ik\mu \Phi,
\ee
with $\mu \equiv \hat\k \cdot \n$, is solved by 
\begin{align} \label{Gamma-sol}
\Gamma(\eta, \k, \q) =&\, e^{ik\mu(\eta_\text{in}-\eta)} \left[\Gamma(\eta_\text{in}, \k, q)  +  \Phi (\eta_\text{in},\k)\right] \nonumber \\ &+2 \int_{\eta_\text{in}}^\eta d\eta' e^{ik\mu(\eta'-\eta)} \frac{\d\Phi(\eta',\k)}{\d\eta'}, 
\end{align}
where $\eta_\text{in}$ is the time when the gravitons were emitted. The last expression will describe the anisotropies in the GW energy spectrum. The first term in (\ref{Gamma-sol}) represents a contribution from the source operator of the Boltzmann equation as an initial condition, its explicit frequency dependence $q$ is a feature of CGWBs. The second term, is the so-called Sachs-Wolfe (SW) effect, and the last one on the integral, corresponds to the integrated Sachs-Wolfe (ISW) contribution.
Anisotropies will emerge when comparing the GW energy density at two different directions in the sky. Therefore, we are interested in computing angular correlations to quantify them. Let us consider a GW arriving at $(\eta_0,\x_0)$, so that, the expansion of (\ref{Gamma-sol}) in spherical harmonics reads 
\begin{align}
&\Gamma_{\ell m}(\eta_0,\x_0,q) = 4\pi (-i)^\ell \int \frac{d^3 \k}{(2\pi)^3} e^{i \k \cdot \x_0}\int_{\eta_\text{in}}^{\eta_0} d\eta \label{Gamma-lm}  \\ &\,  \times \left[ \left[\Gamma(\eta, \k, q)+ \Phi (\eta,\k)\right]\delta(\eta-\eta_\text{in})+ 2 \frac{\d \Phi(\eta,\k)}{\d \eta}\right]\nonumber \\
&\qquad \qquad \qquad \qquad \times Y_{\ell m}^*(\hat \k)j_\ell(k(\eta_0-\eta)), \nonumber  \\
&\, = \Gamma_{\ell m}^0(\eta_0,\x_0,q) + \Gamma_{\ell m}^\text{SW}(\eta_0,\x_0)+\Gamma_{\ell m}^\text{ISW}(\eta_0,\x_0)\label{anisotropies-sum},
\end{align}
where in the last line we have separated and labeled each contribution.

With the full distribution function at hand, we can relate it to the GW energy density as
\begin{align}
  \rho_\text{GW}(\eta,\x) &= \frac{1}{a^4} \int d^3\q \, q f(\eta, \x, \q) \label{eq:f} \\
  &= \rho_c \int d\ln q \int d^2\n\,  \Omega_\text{GW}(\eta, \x, \q), \label{eq:om}   
\end{align}
where $\rho_c$ is the critical energy density of the universe and in the last line, we have used the relation between the GW energy density and energy spectrum.
We can expand the energy spectrum on its homogeneous isotropic background and anisotropic piece as
\be \label{GW-density-spectrum}
\Omega_\text{GW}(\eta,\x,\q) = \bar{\Omega}_\text{GW}(\eta,q)+ \delta \Omega_\text{GW}(\eta,\x,\q).
\ee
The background $\bar{\Omega}_\text{GW}$, often called the monopole, can be obtained through the ensemble average of the energy density spectrum which, by ergodic assumptions, can be computed as the spatial average as follows, relating Eqs. (\ref{eq:f}) and (\ref{eq:om}),
\begin{equation}\label{monopole}
\bar{\Omega}_\text{GW}(\eta,q) = \langle\Omega_\text{GW}(\eta,\x,q) \rangle=\frac{4\pi} {\rho_c}\left(\frac{q}{a(\eta)}\right)^4 \bar{f}(q).
\end{equation}
On the other hand, from the anisotropic piece, we can introduce the GW density contrast,
\begin{align}
\delta_\text{GW}(\eta, \x, \q) &\equiv \frac{\delta \Omega_\text{GW}(\eta,\x,\q)}{\bar{\Omega}_\text{GW}(\eta,q)}\\ 
&= \left[4-n_\text{GW}(q)\right] \Gamma(\eta, \x, \q), \label{gw-contrast}
\end{align}
where we have introduced the GW spectral index, $n_\text{GW}(q) \equiv \d \ln \bar{\Omega}_\text{GW} / \d \ln q$. In the last equality, we show how the GW density contrast is connected to the first-order perturbations of the distribution function.
These are the two main quantities describing the background and anisotropies of the CGWB.

Once we have a way to isolate the CGWB anisotropies, we can describe their statistical properties by computing the angular correlation functions as, 
\begin{align}
    \langle \delta_{\text{GW},\ell m}^{\text{X}}\delta_{\text{GW},\ell' m'}^{*\,\text{Y}}\rangle &= \left[4-n_\text{GW}(q)\right]^{2} \langle \Gamma_{\ell m}^{\text{X}}\Gamma_{\ell' m'}^{*\,\text{Y}}\rangle\\
    &=\delta_{\ell \ell'}\delta_{mm'} C_{\text{GW},\ell}^{\text{XY}}, \label{angular-gw-contrast}
\end{align}
where by direct connection with Eq. (\ref{anisotropies-sum}), $\text{X}$ and $\text{Y}$ represent each of the contributions from the GW anisotropies.
\subsection{Gravitational Waves Induced by Scalar Perturbations}\label{sec:SIGW}

Having established the relation between statistical quantities and GW observables, we will now provide explicit expressions for $\bar\Omega_\text{GW}(q)$ and $\Gamma(\eta_i,\k, q)$, considering that the CGWB is sourced by large primordial fluctuations. We will closely follow and combine the previous computations presented in \cite{Cai:2018dig, Unal:2018yaa, Adshead:2021hnm, Li:2023qua}. 

When perturbing Einstein equations at first order, different spins do not mix. However, due to the non-linearity of Einstein equations, at second order this no longer holds.
In particular, we will focus on the equation of motion for a second-order tensor perturbation $h_{ij}$ sourced by a combination of derivatives of linear scalar perturbations.
To be concrete, let us add a second order, transverse traceless tensor perturbation $h_{ij}$ to the metric (\ref{pert-FLRW}), such that 
\begin{equation*}\label{pert-FLRW+h}
ds^2  = a^2\left[-(1+2\Phi)d\eta^2 + \left[\left(1-2\Phi\right)\delta_{ij}+\frac{h_{ij}}{2}\right]dx^i dx^j\right].
\end{equation*}
After expanding the tensor mode in its two polarizations $\lambda = +,\times$, the second order Einstein equations for the tensor amplitude in Fourier space reads
\be\label{sigw-eom}
h_{\lambda}''(\eta,\q) + 2\mathcal{H}h_{\lambda}'(\eta,\q)+q^2h_{\lambda}(\eta,\q) = \S_{\lambda}(\eta,\q),
\ee
where the source term $\S$ originates from quadratic combinations of the first order scalar perturbation $\Phi$.
In particular, we will focus on scalar perturbations produced when a large primordial curvature perturbation $\zeta$ re-enters the horizon during the radiation-dominated epoch of the universe. In that case, the source is 
\begin{align}
\mathcal{S}_\lambda(\eta,\q)=\int \frac{d^3\p}{(2\pi)^3}& \frac{q^2}{\sqrt{2}}v_\lambda(\theta,\varphi)f\left(|\q-\p|,p,\eta \right) \nonumber \\
&\times \zeta(\p)\zeta(\q-\p),
\end{align}
where 
$\begin{pmatrix} v_{+} \\ v_{\times} \end{pmatrix}= \sin^2\theta \begin{pmatrix} \cos2\varphi \\ \sin 2\varphi \end{pmatrix}$
and we have used that, with our gauge choice, during radiation the relation between the scalar perturbation and the primordial curvature perturbation is $\Phi(\eta,\q) =\frac{2}{x}j_1(x)\zeta(\q)$, so the function $f$ is given by \cite{Cai:2018dig}
\begin{align}
f\left(u,v,x \right) &= 8 \left[j_0(u x) j_0(v x)-2 \frac{j_1(u x) j_0(v x)}{u x}\right. \nonumber \\
& \left.\quad-2 \frac{j_0(u x) j_1(v x)}{v x}+6 \frac{j_1(u x) j_1(v x)}{u v x^2}\right], 
\end{align}
where we have introduced the variables $x=q\eta/\sqrt{3}$, $u =|\q - \p|/q$, $v=p/q$. The large primordial fluctuations we are considering are produced in small scales, so that their two-point function is larger than the near-scale-invariant power spectrum on CMB scales. 
Model builders have devised several mechanisms to produce this feature \cite{Escriva:2022duf}, in this work we will restrict ourselves to the simplest way to capture this behavior, {\it i.e.,} introducing a localized peak at a small-scale $q_*$.
The solution of (\ref{sigw-eom}), obtained through the Green function method, yields
\begin{align} \label{graviton-sol}
h_\lambda(\eta,\q)= 4\int \frac{d^3\p}{(2\pi)^3}& v_\lambda(\theta,\varphi) \mathcal{J}\left(|\q-\p|,p,\eta \right)\nonumber \\ &\times\zeta(\p)\zeta(\q-\p),
\end{align}
where we have introduced the function 
\begin{align}
\mathcal{J}\left(u,v,x \right)=\frac{q}{\eta} \int_0^{\eta}d\eta'\eta'\sin (q(\eta-\eta'))f\left(u,v,x\right).
\end{align}

\subsubsection{Background}
\begin{figure}[t!]
\centering
\begin{tikzpicture}[scale=1.8, transform shape]
\fill[pattern=north east lines] (1,1) rectangle (2,2);
\draw[dashed] (1,1) rectangle (2,2); 
\node at (1.5,1.5) {\Large $\mathcal{O}$};     
\draw[thin] (.5,1.5) -- (1,2) node[pos=0.7, above, sloped] {\small $\zeta_{\p_1-\q_1}$}; 
\draw[thin] (.5,1.5) -- (1,1) node[pos=0.35, below] {$\zeta_{\p_1}$}; 
\draw[thin] (2,2) -- (2.5,1.5) node[pos=0.4, above, sloped]{$\zeta_{\p_2-\q_2}$}; 
\draw[thin] (2,1) -- (2.5,1.5) node[pos=0.9, below, yshift=-1mm]{$\zeta_{\p_2}$}; 
\draw[decorate, decoration={snake, amplitude=0.8mm, segment length=2mm}] (-0.3,1.5) -- (.5,1.5) node[pos=0.2, above, yshift=.3mm] {$h_{\tiny \q_1}^{\lambda_1}$}; 
\draw[decorate, decoration={snake, amplitude=0.8mm, segment length=2mm}]  (3.3,1.5) -- (2.5,1.5) node[pos=0.2, above, yshift=.3mm] {$h_{\tiny \q_2}^{\lambda_2}$}; 
\foreach \x/\y in {1/1, 1/2, 2/1, 2/2, .5/1.5, 2.5/1.5} {
  \fill (\x,\y) circle (.6pt);}
\end{tikzpicture}
\caption{Diagrammatic representation of $\bar{\Omega}_\text{GW}$ according to (\ref{monopole-sigw}) and (\ref{h-two-point}). Each vertex $h\zeta\zeta$ contributes with $4\int \frac{d^3\p_i}{(2\pi)^2}v_{\lambda_i}\mathcal{J}(u_i,v_i,x_i)$, and the dashed box $\mathcal{O}$ indicates connected combinations of $\zeta$'s for point-function expressed in terms of the gaussian perturbation $\zeta_g$ given in Eqs. (\ref{diag-1})-(\ref{diag-3}).}
\label{fig:diagram}
\end{figure}
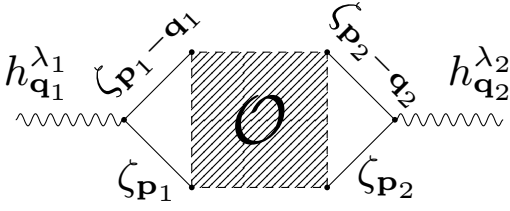

Once the solution (\ref{graviton-sol}) is known, we can compute the background GW energy density spectrum as the monopole term 
\be\label{monopole-sigw}
\bar\Omega_\text{GW}(q) = \frac{q^5}{48\pi^2a^2H^2}
\sum_{\scriptstyle +,\times}\overline{\langle h_{\lambda}(\eta,\q)h_{\lambda'}(\eta,\q')\rangle},
\ee
where the over-bar denotes a time average over the oscillations of the tensor modes. Explicitly the graviton two-point function is given by 
\begin{align} \label{h-two-point}
\langle &h_{\lambda_1}(\eta,\q_1)h_{\lambda_2}(\eta,\q_2)\rangle = 16\int\frac{d^3\p_1}{(2\pi)^3}\int\frac{d^3\p_2}{(2\pi)^3}\\
&\times v_{\lambda_1}(\theta_1,\varphi_1)\mathcal{J}(u_1,v_1,x_1) v_{\lambda_2}(\theta_2,\varphi_2)\mathcal{J}(u_2,v_2,x_2)\nonumber\\ 
&\qquad  \times \left\langle\zeta(\p_1) \zeta(\q_1-\p_1)\zeta(\p_2) \zeta(\q_2-\p_2) \right\rangle. \nonumber 
\end{align}
Since $h\sim\zeta^2$, the background spectrum goes like $\bar{\Omega}_\text{GW} \sim \langle h^2\rangle \sim \langle\zeta^4\rangle$, as shown diagrammatically in Fig. \ref{fig:diagram}.

\begin{figure}[t]
  \centering
    \includegraphics[width=1.1\columnwidth]{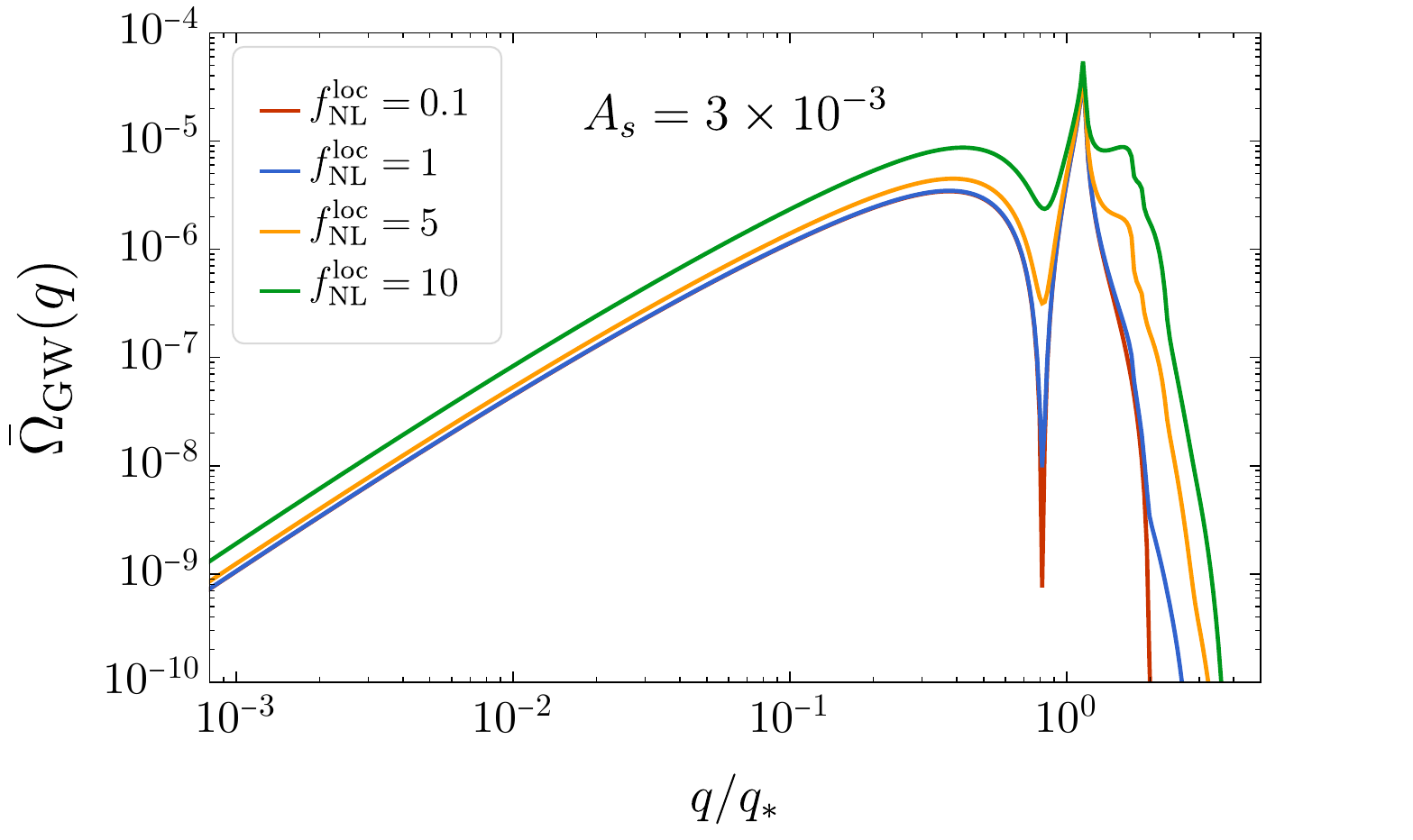}
  \caption{Background induced GW spectrum $\Omega_\text{GW}(q)$ for a monochromatic source with $A_s=3\times10^{-3}$  for various values of $\locfnl$.}
  \label{fig:gwspectrum}
\end{figure}

When $\zeta$ is gaussian, the four-point function can be completely expressed in terms of the two-point one. Nevertheless, as we mentioned in the introduction, we will focus on non-Gaussian primordial fluctuations of the local type (\ref{local-ansatz}). In Fourier space, $\zeta$ is expressed as a combination of a gaussian field $\zeta_g$ as 
\be
\zeta(\q) = \zeta_g(\q) + \frac{3}{5}\locfnl\int \frac{d^3\p}{(2\pi)^3}\zeta_g(\q)\zeta_g(\q-\p), 
\ee
with a primordial power spectrum defined as
\be\label{zeta-power}
\langle \zeta_g(\q)\zeta_g(\q') \rangle = (2\pi)^3\delta^{(3)}(\q+\q')P_{\zeta_g}(q),
\ee
for which the dimensionless power spectrum is $\Delta_{\zeta_g}^2(q) \equiv \frac{q^3}{2\pi^2}P_{\zeta_g}(q)$.
With this expansion, one can decompose the four-point function of $\zeta$ into three different contributions,
$\langle\zeta^4\rangle \sim \langle\zeta_g^4 \rangle + {\locfnl}^2 \langle\zeta_g^6\rangle + {\locfnl}^4 \langle\zeta_g^8\rangle$ and then use Wick's theorem in (\ref{h-two-point}) to express each gaussian correlator in terms of $\langle \zeta_g^2\rangle$. Diagrammatically we have the following, 
\begin{align}
\mathcal{O}\left(\langle \zeta_g^4\rangle \right) &=
\raisebox{2pt}{\begin{tikzpicture}[baseline={(current bounding box.center)}, scale=1, transform shape]
  \draw[dashed] (1,1) rectangle (2,2); 
  \draw[thick, mid arrow] (1,2) -- (2,2) node[pos=0.5, above] {\small $1-\q$};  
  \draw[thick, mid arrow] (2,1) -- (1,1) node[pos=0.5, below] {\small $1$};  
\end{tikzpicture}}\, \subset \bar{\Omega}_\text{GW}^{(0)},\label{diag-1} \\
\mathcal{O}\left(\langle \zeta_g^6\rangle \right) &= 
\raisebox{2pt}{\begin{tikzpicture}[baseline={(current bounding box.center)}, scale=1, transform shape]
  \draw[dashed] (1,1) rectangle (2,2); 
  \foreach \x/\y in {1/1, 2/2} {
    \fill (\x,\y) circle (1.8pt);  
  }
  \draw[thick, mid arrow] (2,2) -- (1,2) node[pos=0.5, above] {\small $1$};
  \draw[thick, mid arrow] (1,1) -- (2,2) node[pos=0.5, below, sloped, yshift=1pt] {\tiny 1+2-$\q$};
  \draw[thick, mid arrow] (2,1) -- (1,1) node[pos=0.5, below] {\small $2$};
\end{tikzpicture}}
+
\raisebox{2pt}{\begin{tikzpicture}[baseline={(current bounding box.center)}, scale=1, transform shape]
  \draw[dashed] (1,1) rectangle (2,2);
  \foreach \x/\y in {1/1, 1/2} {
    \fill (\x,\y) circle (1.8pt);  
  }
  \draw[thick, mid arrow] (1,2) -- (2,2) node[pos=0.5, above] {\small $2-\q$};
  \draw[thick, mid arrow] (1,2) -- (1,1) node[pos=0.5, sloped, above] {\small $1-2$};
  \draw[thick, mid arrow] (2,1) -- (1,1) node[pos=0.5, below] {\small $2$};
\end{tikzpicture}}
+
\raisebox{0pt}{\begin{tikzpicture}[baseline={(current bounding box.center)}, scale=1, transform shape]
  \draw[dashed] (1,1) rectangle (2,2);
  \foreach \x/\y in {1/1, 2/1} {
    \fill (\x,\y) circle (1.8pt);
  }
  \draw[thick, mid arrow] (1,2) -- (2,2) node[pos=0.5, above] {\small $1-\q$};
  \draw[thick, mid arrow] (1,1) to[out=90,in=90] node[pos=0.5, above, sloped] {\small $2-1$} (2,1) ; 
  \draw[thick, mid arrow] (2,1) to[out=-90,in=-90] node[pos=0.5, below, sloped] {\small $2$} (1,1) ;
\end{tikzpicture}}\,\subset \bar{\Omega}_\text{GW}^{(1)},\label{diag-2}\\
\mathcal{O}\left(\langle \zeta_g^8\rangle \right) &= 
\raisebox{2pt}{\begin{tikzpicture}[baseline={(current bounding box.center)}, scale=1, transform shape]
  \draw[dashed] (1,1) rectangle (2,2); 
\foreach \x/\y in {1/1, 1/2, 2/1, 2/2,} {
  \fill (\x,\y) circle (1.8pt);  
}
\draw[thick, mid arrow] (1,2) -- (2,2) node[pos=0.5, above] {\small $3-\q$};
\draw[thick, mid arrow] (1,2) -- (1,1) node[pos=0.5, above, sloped] {\small $1-3$};
\draw[thick, mid arrow] (2,1) -- (1,1) node[pos=0.5, below] {\small $3$};
\draw[thick, mid arrow] (2,1) -- (2,2) node[pos=0.5, above, sloped] {\small $2-3$};
\end{tikzpicture}}
+
\raisebox{2pt}{\begin{tikzpicture}[baseline={(current bounding box.center)}, scale=1, transform shape]
\draw[dashed] (1,1) rectangle (2,2);
\draw[thick, mid arrow] (2,2) -- (1.5,1.5) node[pos=0.7, below, sloped, yshift=1pt] {\tiny 1-3};
\draw[thick] (1,1) -- (1.5,1.5);
\draw[thick, mid arrow] (1.45,1.55) -- (1,2) node[pos=0.3, below, sloped, yshift=0.7pt] {\tiny 2-3};
\draw[thick] (1.55,1.45) -- (2,1);
\draw[thick, mid arrow] (1,2) -- (2,2) node[pos=0.5, above] {\small 1-$\q$-2-3};
\draw[thick, mid arrow] (2,1) -- (1,1) node[pos=0.5, below] {\small 3};
\foreach \x/\y in {1/1, 1/2, 2/1, 2/2} {
  \fill (\x,\y) circle (1.8pt);}
\end{tikzpicture}}
+
\raisebox{2pt}{\begin{tikzpicture}[baseline={(current bounding box.center)}, scale=1, transform shape, ]
  \draw[dashed] (1,1) rectangle (2,2);
  \foreach \x/\y in {1/1, 1/2, 2/1, 2/2} {
    \fill (\x,\y) circle (1.8pt);
  }
  \draw[thick, mid arrow] (1,2) to[out=90,in=90] node[pos=0.5, above, sloped] {\small 1+2-$\q$} (2,2); 
  \draw[thick, mid arrow] (2,2) to[out=-90,in=-90] node[pos=0.5, above, sloped, yshift=-2pt] {\small 3}(1,2); 
  \draw[thick, mid arrow] (1,1) to[out=90,in=90] node[pos=0.5, below, sloped, yshift=2pt] {\small 2-1} (2,1); 
  \draw[thick, mid arrow] (2,1) to[out=-90,in=-90] node[pos=0.5, below, sloped] {\small 2} (1,1);
\end{tikzpicture}}\,\subset \bar{\Omega}_\text{GW}^{(2)}\label{diag-3}.
\end{align}
Each thick black-line with an arrow contributes with a $\Delta_{\zeta_g}^2(k)$ evaluated at the momentum configuration indicated above or below the arrow. The numbers represent the subindex of the internal momentum $\p_i$, for instance $1-2$ corresponds to $\p_1-\p_2$ and so on. Each black dot corresponds to a factor of $3/5 \locfnl$, all this entering in (\ref{h-two-point}) (See \cite{Adshead:2021hnm}). 
Therefore, the background energy spectrum can be represented as 
\be \label{GW-spectrum-fnl}
\bar{\Omega}_\text{GW} = \bar{\Omega}_\text{GW}^{(0)} + \bar{\Omega}_\text{GW}^{(1)} + \bar{\Omega}_\text{GW}^{(2)},
\ee
where the superscript ${(n)}$ in each contribution indicates the power of $\left({\locfnl}^{2n}\right)$ to which is proportional. Notice that (\ref{GW-spectrum-fnl}) depends on even powers of $\locfnl$, therefore, it is insensitive to its sign.
Since we are interested in GWs produced at small scales, in principle we can approximate the full Gaussian distribution dominated by the power spectrum at these scales by
\be\label{PS-short}
\Delta_{\zeta_g}^2(q) \simeq A_{s}q_{*} \, \delta(q-q_*).
\ee
The GW production is a local process, where from an EFT perspective, we only consider the relevant scales involved, in this case $q_*^{-1}$. The peak frequency also sets the initial time in (\ref{Gamma-sol}) as $\eta_\text{in}\sim 1/q_*$. 
The corresponding background GW spectrum is depicted\footnote{The spectrum was computed with the methods of Ref. \cite{Adshead:2021hnm} publicly available at \url{https://github.com/zachjweiner/ngsigw-results}.} in Fig.\ref{fig:gwspectrum}. 

By showing the spectrum as a function of the peak normalized frequency, $q/q_*$, allows us later to accommodate the frequency peak within the sensitivity curve of any survey of interest. In particular, in Sec. \ref{sec-V}, we will consider the observational prospects for anisotropies when the signal is located in the sensitivity region of the combination of the Einstein Telescope (ET) and the Cosmic Explorer (CE).

\subsubsection{Anisotropies}

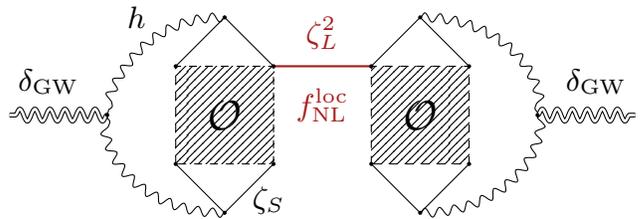
\begin{figure}[t]
\centering
\begin{tikzpicture}[scale=1.3, transform shape]

\node at (0.6,2.5) {$h$};  
\draw[dashed] (1,1) rectangle (2,2); 
\fill[pattern=north east lines] (1,1) rectangle (2,2);
\node at (1.5,1.5) {\large $\mathcal{O}$}; 
\draw[dashed] (3,1) rectangle (4,2);
\fill[pattern=north east lines] (3,1) rectangle (4,2);
\node at (3.5,1.5) {\large $\mathcal{O}$}; 
\draw[thin] (1,2) -- (1.5,2.5); 
\draw[thin] (2,2) -- (1.5,2.5); 
\draw[thin] (1,1) -- (1.5,0.5); 
\draw[thin] (2,1) -- (1.5,0.5)node[pos=0.1, below, yshift=-.7mm] {$\zeta_S$}; 
\draw[decorate, decoration={snake, amplitude=0.6mm, segment length=2mm}] (1.5,2.5) to[out=180,in=75] (0.3,1.5); 
\draw[decorate, decoration={snake, amplitude=0.6mm, segment length=2mm}] (1.5,0.5) to[out=180,in=-90] (0.3,1.5); 
\draw[decorate, 
      decoration={snake, amplitude=0.8mm, segment length=2mm},
      double, 
      double distance=1pt] 
      (0.3,1.5) -- (-0.7,1.5) node[pos=0.6, above, yshift=1mm] {$\delta_\text{GW}$};
\draw[thin] (3,2) -- (3.5,2.5); 
\draw[thin] (4,2) -- (3.5,2.5); 
\draw[thin] (3,1) -- (3.5,0.5); 
\draw[thin] (4,1) -- (3.5,0.5); 
\draw[decorate, decoration={snake, amplitude=0.6mm, segment length=2mm}] (3.5,2.5) to[out=0,in=90] (4.7,1.5); 
\draw[decorate, decoration={snake, amplitude=0.6mm, segment length=2mm}] (3.5,0.5) to[out=0,in=-90] (4.7,1.5); 
\draw[decorate, 
      decoration={snake, amplitude=0.8mm, segment length=2mm},
      double, 
      double distance=1pt] 
      (4.7,1.5) -- (5.7,1.5) node[pos=0.6, above, yshift=1mm] {$\delta_\text{GW}$};
\draw[carnelian, thick] (2,2) -- (3,2) node[pos=0.5, above, yshift=.5mm] {$\zeta_L^2$} node[pos=0.5, below, yshift=-1mm] {$\locfnl$};; 
\foreach \x/\y in {1/1, 1/2, 2/1, 2/2, 3/1, 3/2, 4/1, 4/2, 1.5/2.5, 1.5/0.5, 3.5/2.5, 3.5/0.5, 4.7/1.5, 0.3/1.5} {
  \fill (\x,\y) circle (.6pt);}
\end{tikzpicture}
\caption{Diagrammatic representation of $\langle \delta_\text{GW}^2\rangle$ built from the diagram of Fig.\ref{fig:diagram}. Thanks to $\locfnl$, a long-wavelength mode $\zeta_L$ is able to connect both sides of the figure, this is the so-called $\locfnl$ bridge.}
\label{fig:delta-diagram}
\end{figure}

As we mentioned at the beginning of this section, anisotropies in the GW energy spectrum can arise from both initial conditions and propagation effects.
The former contribution, for the case of SIGWs, appears only when there is a coupling between the short scale $q_*^{-1}$ at which the GW was produced and a cosmological scale $k^{-1}$. Primordial local non-Gaussianity enables this coupling \cite{Bartolo:2019zvb, Tada:2015noa}. 
In the absence of short-long wavelength couplings, different patches in the sky ---of the size of the horizon at the GW production--- would look the same.
Whereas, if a long wavelength mode is able to influence the local physics of the GW production at scales $q_*^{-1}$, those patches would no longer exhibit the same energy density, therefore one expects to detect such a difference as an anisotropy at large scales $k^{-1}$. Technically, this implies that the correlation of the energy spectrum at two different locations in the sky is due only to a long-wavelength mode, as shown diagrammatically in Fig. \ref{fig:delta-diagram}.

Let us quantify the initial condition anisotropy due to the previous effect. We can split the Gaussian component of the primordial curvature perturbation as the sum of uncorrelated short- and long-wavelength pieces
\be
\zeta_g= \zeta_S(\q)+ \zeta_L(\k), 
\ee
with $\q\gg\k$, such that $\Delta_{\zeta_g}^{2} = \Delta_S^{2}+\Delta_L^{2}$.
For long-wavelength modes, $\Delta_L^2(k)$ is the nearly scale-invariant power spectrum measured at CMB scales, which we will approximate as $\Delta_L^2(k) = A_L \sim 10^{-9}$. While $\Delta_S^2(q)$, associated to the GW production at short scales $q_*^{-1}$ will be given by (\ref{PS-short}). Considering the small amplitude of the scale-invariant power spectrum, we will compute the angular two-point function of the anisotropies up to linear order in $\Delta_L^2$, as shown with solid red line connecting the center of the diagram in Fig. \ref{fig:delta-diagram}. As a consequence, when computing $\langle\delta_\text{GW}^2\rangle\sim\langle h^4 \rangle\sim \langle \zeta^8 \rangle$, the allowed operators producing anisotropies linear in $\Delta_L^2$, are the same ones that contribute to $\bar{\Omega}_\text{GW}^{(0)}$ and $\bar{\Omega}_\text{GW}^{(2)}$ with additional symmetry factors given the topology of the diagram in Fig. \ref{fig:delta-diagram} (see \cite{Li:2023xtl}).
Concretely, when $\locfnl$ is present, one can introduce an energy density spectrum associated to the diagrams contributing linearly in $\Delta_L^2$ to the anisotropies as
\be
\Omega_\text{NG}(\eta_\text{in},q) = 2^3\bar{\Omega}_\text{GW}^{(0)}(\eta_\text{in},q) +2^2\bar{\Omega}_\text{GW}^{(1)}(\eta_\text{in},q). 
\ee
Therefore, the GW density contrast associated to the initial condition reads \cite{Li:2023qua}
\be\label{initial-anisotropy}
\delta_\text{GW}^0(\eta_\text{in},\k,q) = \frac{3}{5}\locfnl \frac{\Omega_\text{NG}(\eta_\text{in},q)}{\bar{\Omega}_\text{GW}(\eta_\text{in},q)}\zeta_L(\k). 
\ee

For the anisotropies due to propagation effects, these are generated when the long-modes of interest re-entered the horizon during the matter-dominated (MD) universe. The evolution of the scalar perturbation $\Phi$ is related to $\zeta$ through
\be \label{transfer-matt}
\Phi(\eta,\k) = \frac{3}{5}T(k)g(\eta)\zeta_L(\k),
\ee
where $T(k)$ is the transfer function and $g(\eta)$ is the growth factor. For long-wavelength modes, one can use $T(k\ll1)=1$ while for short ones analytical expressions are often difficult to obtain but already implemented in the numerical computations. For the case of the SW effect, one needs to consider $g(\eta_\text{in})=1$, while for the ISW, we have to take into account the evolution of the growth factor $g(\eta)$ until our present epoch. More concretely, the relevant quantity from the ISW entering in the computation of the angular correlations is
\be\label{transfer-ISW}
I_\ell^\text{ISW}(\eta_0,\eta_\text{in},k) = \frac{6}{5} \int_{\eta_\text{in}}^{\eta_0}d\eta \frac{dg(\eta)}{d\eta} j_\ell(k(\eta_0-\eta)).
\ee
The impact of the last expression on the different correlations strongly depends on the growth rate $dg/d\eta$. Roughly, the growth rate has large contributions at two moments during the cosmic evolution. The first one appears in the RD universe, at short scales, leading to the {\it early} ISW effect, while the second one, described in the introduction, appears in the $\Lambda$D universe and leads to the {\it late} ISW effect. As we previously stressed, the long-wavelength modes of our interest re-entered the horizon during MD, therefore, we expect contributions from the {\it late} ISW (late-ISW). Concretely, we need to consider the lower limit of the integral in (\ref{transfer-ISW}) at most as the time of the MD-$\Lambda$D equivalence, $\eta_{\Lambda}$.
On the other hand, when the universe is dominated by the cosmological constant, one has the well-known approximation for the growth-rate \cite{Maggiore:2018sht, Dodelson:2020bqr, Lyth:2009imm} 
\be
\frac{d\ln g}{d\ln a} = \Omega_\text{m}^\gamma(a) - 1, 
\ee
with $\Omega_\text{m}(a) = \Omega_\text{m}/(\Omega_\text{m}+(1-\Omega_\text{m})a^3)$ and $\gamma \simeq 0.55 \text{-} 0.6$, and $\Omega_\text{m}$ the matter density parameter. In terms of the conformal time, one finds that the growth-rate is well described by
\be
\frac{dg}{d\eta} = -\frac{4}{5}\left(\frac{\eta}{\eta_0}\right)^3,
\ee
then, the exact solution of (\ref{transfer-ISW}) reads 
\begin{align}\label{isw-transfer-I}
I&_\ell^\text{ISW}=\frac{12\sqrt{\pi}}{25 \cdot 2^{\ell}}\frac{(\eta_0-\eta_\Lambda)^{\ell+1}k^{\ell}}{\Gamma(3/2+\ell)} \sum_{n=0}^{3}\binom{3}{n}\frac{\left(\frac{\eta_\Lambda}{\eta_0}-1\right)^n}{\ell+1+n} \\&\times\,_1F_2\left(\frac{l+1+n}{2}; \frac{3}{2}+l,\frac{l+3+n}{2};-\frac{k^2(\eta_0-\eta_\Lambda)}{4}\right), \nonumber 
\end{align}
where $\,_1F_2(a;b,c;d)$ is a generalized hypergeometric function. Therefore, according to (\ref{angular-gw-contrast}), the total autocorrelation of GW anisotropies is given by \cite{Li:2023qua}
\begin{align}\label{eq:Cl_gwgw}
C_\ell^\text{CGW} &=\frac{18\pi\Delta^2_{L}}{25\ell(\ell+1)}\left[\frac{\Omega_\text{NG}(q)}{\bar{\Omega}_\text{GW}(q)}\locfnl + (4-n_\text{GW}(q)) \right]^2 \nonumber \\ &\quad + C_\ell^\text{ISW}, 
\end{align}
where the first term come from the initial condition and SW anisotropies, and last one, coming from the ISW is given by 
\begin{align} \label{eq:Cl_isw-isw}
\frac{C_\ell^\text{ISW}}{\left[4-n_\text{GW}(q)\right]^2}=&\,\Delta^2_{L}\int_0^\infty \frac{dk}{k}T^2(k) \left[I_\ell^\text{ISW}(\eta_0,\eta_\Lambda,k)\right]^2.
\end{align}

\begin{figure}[]
    \includegraphics[width=\columnwidth]{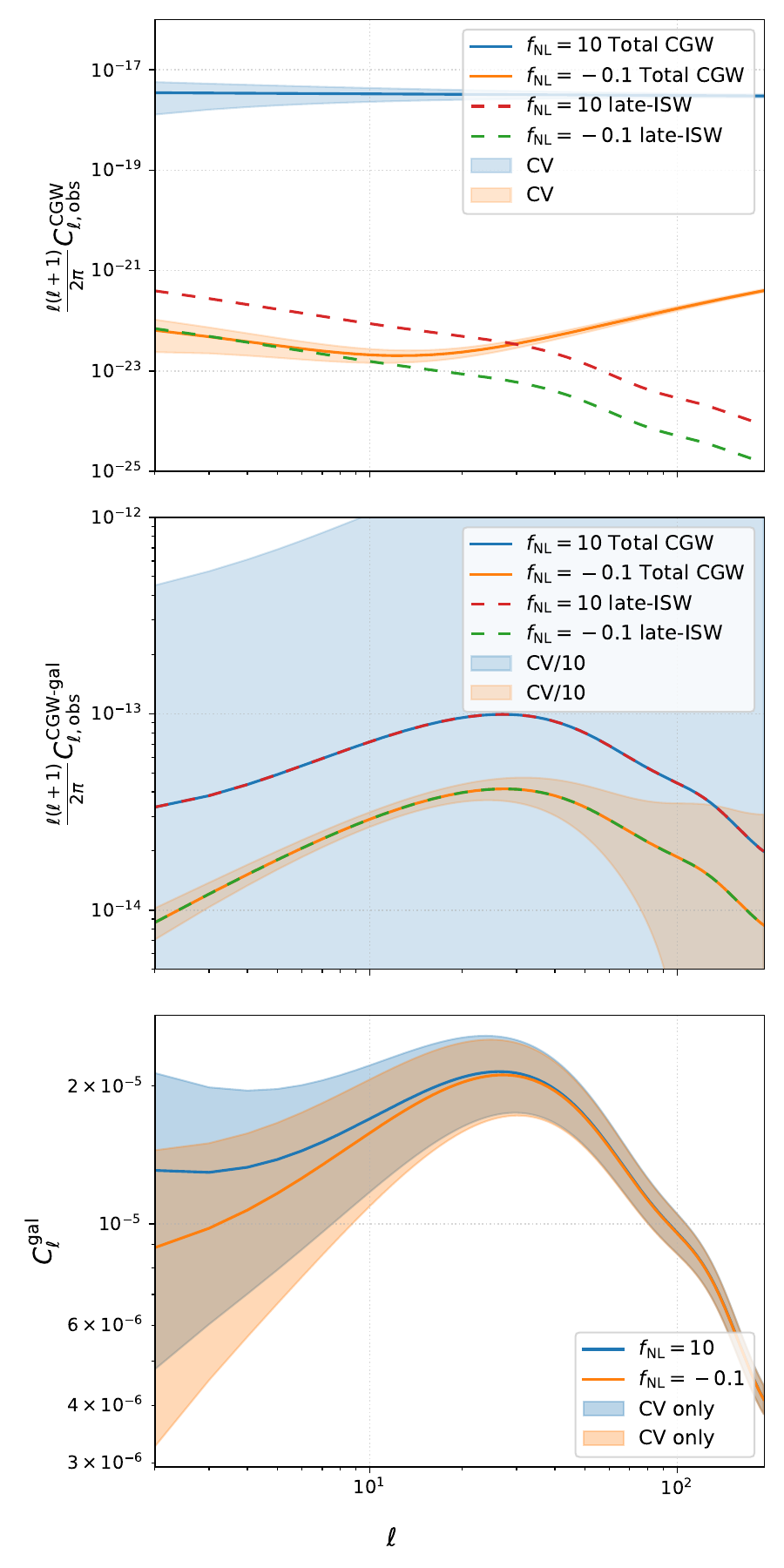}
    \caption{Comparison between the multiple correlations $C_{\ell}^{XY}$ used in this work. For all panels, blue line corresponds to $f_{\rm NL}=10$ and orange line to $f_{\rm NL}=-0.1$. The GW contributions are multiplied by the monopole evaluated at $f_{\text{pivot}}=63\text{Hz}$. The shaded regions correspond to $1\sigma$, obtained from the variance computed in Eq. (\ref{eq:variance}) considering only cosmic variance (CV) and ignoring detector noise. 
    \textbf{Top panel:} Angular power spectrum for the CGWB anistropies sourced by SIGW, as presented in Section \ref{sec-II}. The red and green dashed lines present the correlation of the late-ISW component of the CGWB anisotropies as described by Eq.(\ref{eq:Cl_isw-isw}).
    \textbf{Middle panel:} Main result of our work. Angular power spectrum for the cross-correlation between galaxies and CGWB anisotropies coming from SIGW using all the contributions. Similarly as the top panel, the red and green dashed lines are the cross-correlation when using only the late-ISW component. The computation is presented in Section \ref{sec-IV}.
    \textbf{Bottom panel:} Angular galaxy power spectrum for the third redshift bin of a LSST-like survey as described in Section \ref{sec-III}.}
    \label{fig:cls_comparison}
\end{figure}

In the top panel of Fig.\ref{fig:cls_comparison}, we show\footnote{The angular power spectra were computed using modifications of \texttt{GW\_CLASS} \cite{Schulze:2023ich} and \texttt{Multi\_CLASS} \cite{Bellomo:2020pnw, Bernal:2020pwq, Diego_Blas_2011}.} the angular power spectrum for the autocorrelation of SIGW anisotropies for different values of $\locfnl$.
The blue and orange lines are computed using all the components of the GW anisotropies, as described in Eq.(\ref{anisotropies-sum}). 
The green line shows how this power spectrum changes when we only consider the late-ISW component.
Notice that, the different power spectra are computed using full numerical integration over redshift and momenta, that is, without relying in the commonly used Limber approximation that fails in small $\ell$ \cite{Bernal:2020pwq}.

\section{Tracers of LSS}\label{sec-III}
In the following, we will consider the galaxy density contrast as our LSS tracer. 
The observed galaxy density contrast in a given direction $\n$ in the sky is given by \cite{Giannantonio:2008zi}
\be \label{deltag}
\delta_g (\n) = \int dz \, b(z,k) \frac{dN(z)}{dz} \delta_\text{m}(\n, z),
\ee
where $b(z,k)=b_g(z)+\Delta b(z,k)$ is the galaxy halo bias relating visible and dark matter, where $b_g$ is the linear Gaussian bias and $\Delta b(z,k)$ corresponds to the non-Gaussian one, $dN/dz$ is the so called selection function, and $\delta_\text{m}$ is the matter density perturbation, which is related to the primordial perturbations $\zeta$ as,
\be \label{matter-zeta} 
\delta_\text{m}(\k, z) = \frac{2}{5}\frac{k^2}{\Omega_\text{m}H_0}T(k)D(z)\zeta_L(\k). 
\ee

We assume that galaxies are obtained from a LSST-like survey \cite{LSSTDarkEnergyScience:2018jkl}. For photometric surveys, the galaxy sample is split in redshift bins due to uncertainties in the determination of the exact redshift of galaxies. Because of this, each redshift bin has its own selection function $dN/dz$ given by the statistical distribution of the estimated redshift of each galaxy \cite{DeVicente:2015kyp}.
In particular, in our case we consider that the redshift selection functions are five normalized Gaussian redshift bins of the form 
\be \label{galaxy-selection}
\frac{dN}{dz} = \frac{1}{\sqrt{2\pi\sigma_z^2}}\exp{\left[-\frac{(z-z_\text{bin})^2}{2\pi\sigma_z^2}\right]},
\ee
centered at redshifts $z_{\rm bin}=[0.3, 0.5, 0.7, 0.9, 1.1]$ each with a width of $\sigma_{z}=0.1$.
We also consider that each redshift bin has its correspondent Gaussian galaxy bias, estimated by the relation $b_g(z) = 0.95/D(z)$, where $D(z)$ is the normalized linear density growth factor with $D(z=0)=1$ \cite{LSSTDarkEnergyScience:2018jkl}.
This gives us the following values for the galaxy bias $b_{g}=[1.56, 1.73, 1.91, 2.10, 2.29]$.

After expanding (\ref{deltag}) in spherical harmonics as 
\begin{align}\label{angular-deltag}
\delta_{\text{g},\ell m} = 4\pi (-i)^\ell &\int \frac{d^3 \k}{(2\pi)^3} \int dz \, b(k,z) \frac{dN(z)}{dz} \nonumber  \\
& \times \delta_\text{m}(\k, z) j_\ell(k \chi(z)) Y_{\ell m}^*(\hat \k), 
\end{align}
we can compute the galaxy angular power spectrum (APS) as,
\begin{align} \label{eq:cl_galgal}
\langle \delta_{\text{g},\ell m}\delta_{\text{g},\ell' m'}^{*}\rangle &= \delta_{\ell \ell'}\delta_{mm'}C_{\ell}^\text{gal}\\
C_{\ell}^\text{gal}&= 4\pi \int_0^{\infty} \frac{dk}{k}\Delta^2(k) I_{\ell}^{\text{gal}}(z_1,k)I_{\ell}^{\text{gal}}(z_2,k),
\end{align}
where $\Delta^2(k)\equiv (k^3/2\pi^2) P(k)$ is the dimensionless matter power spectrum at $z=0$, with $P(k)$ given by 
\be\label{matter-ps}
\left\langle \delta_\text{m}(\k_1)\delta_\text{m}^*(\k_2) \right\rangle = (2\pi)^3\delta^{(3)}(\k_1-\k_2)P(k_1)
\ee
and we have defined
\be \label{eq:I_g}
I_{\ell}^\text{gal}(z,\k) \equiv \int dz \, b(z,k) \frac{dN(z)}{dz}\, D(z) j_{\ell}(k \chi(z)),
\ee
where $\chi(z) = \eta_0 - \eta(z)\equiv \eta_0- \eta_z$ is the comoving distance. 

\subsection{Scale-dependent bias}

When the gravitational potential deviates from the Gaussian case, as (\ref{local-ansatz}), the way dark matter collapses into halos changes, such that the galaxy bias acquires a scale dependence \cite{Dalal:2007cu, Slosar:2008hx}.
We can write the scale-dependent bias due to $\locfnl$ as follows \cite{Bernal:2020pwq},
\begin{equation} \label{eq:scale_dep}
    \Delta b(z,k) = 3\locfnl \frac{\Omega_\text{m}H_0^2}{k^{2}T(k)D(z)}(b_{g}-p)\delta_{c},
\end{equation}
where $b_g$ is the constant Gaussian bias and $\delta_{c}=1.686$ is the critical value of spherical collapse for halo formation \cite{Bellomo:2020pnw}. We note that throughout the text we have set the speed of light, $c=1$. For the numerical calculation, we set $c=3\times10^{8} \text{m/s}$.
For our forecasts, we also fix $p=1$, which is customary in many analyses and is considered the prediction for a mass-selected galaxy/halo sample \cite{Riquelme:2022amo}.

In the bottom panel of Figure \ref{fig:cls_comparison}, we show the galaxy angular power spectrum including the scale-dependent bias for different values of $\locfnl$. The shaded regions correspond to $1\sigma$ computed using only cosmic variance (CV) and ignoring shot-noise.

\section{GW$\times$LSS}\label{sec-IV}
Having settled the ground, in the following we will compute the angular cross-correlation between the anisotropies of the CGWB and the galaxy density contrast, defined as
\be\label{cross-correlation}
\langle \delta_{\text{GW},\ell m}\delta_{\text{gal},\ell' m'}^{*}\rangle =\delta_{\ell\ell'}\delta_{mm'} C_{\ell}^\text{CGW-gal}.
\ee
In order to gain some physical intuition let us first estimate the result considering a few approximations. For the galaxy density-contrast (\ref{deltag}) we will consider the Dirac delta limit for the selection function (\ref{galaxy-selection}), $dN/dz = \delta(z-z_\text{bin})$, so we save doing the redshift integral. Additionally, we will fix the transfer function $T(k)=1$\footnote{This approximation is valid up to $\ell_\text{max}\approx 25$ for $k_\text{eq}=0.01 \text{Mpc}^{-1}$ and  $z = 0.7$} when expressing the matter density contrast in terms of the primordial curvature perturbation as in Eq. (\ref{matter-zeta}). Therefore, at the moment of computing the correlation, galaxies contribute to the integral  in $k$ with a term of the form: $\delta_\text{gal}\sim (k^2 + 1)j_\ell(\eta_0-\eta_z)$, where the quadratic term in $k$ comes from the linear contribution of the galaxy bias, the constant one comes from the scale dependent bias (\ref{eq:scale_dep}) and the spherical Bessel function it is evaluated at $\chi(z_\text{bin})$. On the other hand, for the GW anisotropies we have 
\be \label{delta-gw-separated}
\delta_\text{GW}(\eta_\text{in},\eta_\Lambda) = \delta_\text{GW}^{0}(\eta_\text{in})+ \delta_\text{GW}^\text{SW}(\eta_\text{in})+ \delta_\text{GW}^\text{ISW}(\eta_\Lambda), 
\ee
where we have separated the terms that depends on the differences $(\eta_0-\eta_\text{in})$ and  $(\eta_0-\eta_\Lambda)$ only indicating the last term, and we have omitted the spherical harmonics indices. Let us first analyze the cross correlation, considering only the the first two terms, which are given by the sum of (\ref{initial-anisotropy}) and (\ref{transfer-matt}) evaluated at $\eta=\eta_\text{in}$ and multiplied by the spherical Bessel. Since these terms do not have any dependence on $k$ besides the one that comes from the primordial curvature perturbation, then, they contribute to the integral in $k$ with a term of the form $\delta_\text{GW}(\eta_\text{in})\sim j_\ell(k(\eta_0-\eta_\text{in}))$. To sum up, the correlation (\ref{cross-correlation}) for the aforementioned terms, depends on the computation of 
\begin{align}
I_\ell^{(\pm)} = \int_{0}^{\infty}dk k^{\pm 1} j_\ell(k(\eta_0-\eta_\text{in}))j_\ell(k(\eta_0-\eta_z)),
\end{align}
where the indices $+$ and $-$ indicate they come from the linear and scale-dependent bias terms respectively and we have the hierarchy $\eta_0 \gtrsim \eta_z \gg \eta_\text{in}$. The exact solution of the previous integrals are given by a combination of Gamma and Gaussian hypergeometric functions, and we can express them as 
\begin{align}
I_\ell^{(\pm)} \sim \left( \frac{\eta_0-\eta_z}{(\eta_0-\eta_\text{in})^{(3\pm1)/2}}\right)^\ell \frac{\ell^{\pm 1/2}}{(\ell+1/2)},
\end{align}
Given the difference in the conformal times of the galaxy observation $\eta_z$ and the GW emission time, $\eta_\text{in}$ the cross correlation is suppressed for those terms even for small $\ell$. Therefore, we expect they do not contribute to it as we show in the full numerical computation\footnote{We consider the following fiducial cosmology $h=0.678$, $\Omega_{b}=0.0487$, $\Omega_{\rm cdm}=0.2612$, $\Omega_{k}=0$, amplitude of the scale-invariant power spectrum, $A_{L}=2.100549\times10^{-9}$, $k_{\rm pivot}=0.05$, $n_{s}=0.96605$, and setting $f_{\rm dec}(\eta_{\rm ini})=0$, considered fiducial for \texttt{GW\_CLASS} \cite{Schulze:2023ich}. The angular power spectra are computed from $\ell_{\rm min}=2$ up to $\ell_{\rm max}=192$, which is expected to be the smallest scale for GW anisotropies in most GW detectors \cite{Alonso:2020rar}} displayed in the middle panel of Fig. \ref{fig:cls_comparison}. By looking at the green-dashed curve, we can see that the correlation is enhanced on large scales and completely driven by the ISW effect, then,  $C_{\ell}^\text{CGW-gal}\simeq C_\ell^\text{ISW-gal}$.
Thus, the information coming from the initial condition (\ref{initial-anisotropy}) and the SW are subdominant when combining the GW and LSS data. This tells us that when we omit the GW background terms in the cross-correlation then we cannot distinguish between the different cosmological sources.
Thus, \textbf{our result is general for all the possible sources of CGWB anisotropies.}
Having said this, let us analytically estimate the behavior of the ISW galaxy density contrast angular cross-correlation, 
\be \label{angular-isw-gal}
\langle \Gamma_{\ell m}^\text{ISW}\delta_{g, \ell m}^{*}\rangle= \delta_{\ell\ell'}\delta_{mm'} C_\ell^\text{ISW-gal},
\ee  
which is given by the computation of 
\be \label{cl-isw-gal} 
C_\ell^\text{ISW-gal} = \int_0^{\infty} \frac{dk}{k}\Delta_L^2(k) I_{\ell}^\text{ISW}(\eta_\Lambda,k)I_{\ell}^\text{gal}(\eta_z,k)T^2(k)
\ee
where the terms inside the integral are given by (\ref{isw-transfer-I}) and (\ref{eq:I_g}). For the ISW part, we will consider $n=0$ in (\ref{isw-transfer-I}), such that the relevant computation is given by 
\begin{align}
I_\ell^{(\pm)} &= \int_{0}^{\infty} dk\, k^{\ell\pm1} j_\ell(k (\eta_0-\eta_z)) \\ 
& \times \,_1F_2\left(\frac{l+1}{2}; \frac{3}{2}+l,\frac{l+3}{2};-\frac{k^2(\eta_0-\eta_\Lambda)}{4}\right), \nonumber
\end{align}
where the exact solutions are
\begin{align}
&I_\ell^{(\pm)}=\frac{2^{\ell-(1\mp1)/2}}{(\eta_0-\eta_z)^{\ell+1\pm1}}\Gamma\left(\ell+\frac{1\pm1}{2}\right) \\
&\times \,_3F_2\left(\pm\frac{1}{2},\frac{1+\ell}{2},\ell+\frac{1\pm1}{2}; \frac{3}{2}+\ell, \frac{3+\ell}{2}; \frac{\eta_0-\eta_\Lambda}{(\eta_0-\eta_z)^2}\right) \nonumber
\end{align}
By expanding the generalized hypergeometric function $\,_3F_2$ keeping the leading term and using the approximation $\frac{\Gamma(\ell)}{\Gamma(3/2+\ell)} \simeq \frac{1}{\ell \sqrt{(\ell+1)}}$, we estimate the cross-correlation (\ref{cl-isw-gal}) as 
\begin{align}
C_\ell^\text{ISW-gal} &\sim \frac{24\sqrt{\pi}}{50}\frac{\Delta_L^2(k) (\eta_0-\eta_z)}{\ell \sqrt{(\ell+1)^3}}\times  \\
&\left(\frac{\ell\, D(z_*)b_g(z_*)}{\Omega_\text{m}(\eta_0-\eta_z)^2H_0} + \frac{3}{2}(b_g(z_*)-1)\delta_cH_0^2\locfnl \right),\nonumber
\end{align}
where we have assumed $\eta_z=\eta_\Lambda$, and given our initial considerations, it describes the behavior of the curves in the middle panel of Fig. \ref{fig:cls_comparison} up to $\ell \sim 25$.

From the analytical estimation, we can appreciate that the behavior of the cross-correlation of LSS and the CGWB anisotropies differs from $C_{\ell}^{\text{AGW-gal}}\propto (\ell + 1/2)^{-1}$ obtained from the cross-correlation of LSS with the astrophysical gravitational wave background (AGWB) anisotropies \cite{Alonso:2020mva}. Our results indicate that measuring the cross-correlation with LSS could serve as a tool to distinguish between the sources of SGWB anisotropies.

In addition, to study the amplitude of the correlation against the noise of the detectors and the cosmic variance, we must include the effects of the background through
\be \label{eq:cl_ggw}
C_{\ell, \text{obs}}^\text{CGW-gal} \equiv \bar{\Omega}_\text{GW}(q)(4-n_\text{GW}(q)) C_\ell^\text{ISW-gal}. 
\ee
For the particular case of non-Gaussian SIGW, something interesting happens. As presented in Section \ref{sec:SIGW}, $\locfnl$ enters the background of the CGWB. Since the observational quantity of interest is $C_{\ell, \text{obs}}^\text{CGW-gal}$ , this implies that, for the cross-correlation, the GW anisotropies still include a $\locfnl$ fromm the background. This is illustrated in the middle panel of Figure \ref{fig:cls_comparison} where the overall amplitude of the cross-correlation increases as we increase $\locfnl$.
On the other hand, the LSS component of the cross-correlation also induces an effect due to $\locfnl$ in the scale-dependent bias. This can be seen on large scales, where, around $\ell=10$, the tilt of $C_{\ell, \text{obs}}^\text{CGW-gal}$ becomes more pronounced.
These results highlight the importance of $\locfnl$ in the cross-correlation of CGWB anisotropies and LSS.

\section{Detection prospects and Forecast}\label{sec-V}

In this section, we will assess the detectability of the cross-correlation signal by performing an SNR analysis considering an optimistic detection threshold of $\text{SNR}>1$. We begin by estimating the maximum SNR that could come from the ISW effect. Next, we compute the SNR of the cross-correlation for a LSST-like survey. Lastly, after studying the detectability of the CGWB-LSS cross-correlation signal, we use it in a Fisher-matrix forecast of the constraints on $f_{\rm NL}^{\rm loc}$.\footnote{Throughout this section we will be using $f_{\rm NL}^{\rm loc}$ and $f_{\rm NL}$ interchangeably.}

Our main focus is the detectability of the CGWB-LSS cross-correlation limited by cosmic variance. 
As an extension, we also include detector noise in the GW anisotropies.
The noise of the GW anisotropies used in this section, $N_\ell^\text{GW}$, is estimated for the combination of the Einstein Telescope (ET) and the Cosmic Explorer (CE) using the \texttt{schNell} code \cite{Alonso:2020rar}. This code computes the map-level noise of a GW detector as a function of frequency, instrumental response, and geometric configuration. 
We adopt an optimistic scenario where the amplitude of the monopole signal is approximately 5 orders of magnitude above the sensitivity of the GW detector for a fixed pivot frequency, $f_{\rm pivot}=63\rm{Hz}$. 

We remark that our result on the presence of the cross-correlation is independent of the fiducial parameters of the gravitational background and the parameters to generate anisotropies.
This implies that the GW anisotropies model could be modified to match different GW detectors, by adjusting the pivot frequency and amplitude, and our results will still hold.
For example, we could have computed this cross-correlation for other GW detectors such as LISA \cite{LISACosmologyWorkingGroup:2022kbp} by adjusting the pivot frequency in our model and computing its respective noise. Instead, we chose the ET+CE combination in particular because it has better sensitivity to a larger number of multipoles than LISA on large scales \cite{Alonso:2020rar}.

As discussed in Section \ref{sec-IV}, the presence of the cross-correlation signal, driven by the late-ISW effect, is general for all possible sources of CGWB anisotropies.
To investigate how the SNR changes between sources, we will compare the detectability of the cross-correlation for two CGWB anisotropy scenarios.
In the first one, we focus on SIGW with $\locfnl$ in both anisotropies and the background.
In the second scenario, we study the detectability for general CGWB anisotropies in which the initial contribution from PNG does not appear. For this case, we model the cross-correlation ignoring the initial GW contribution due to PNG by setting $f_{\rm NL}^{\rm GW}=0$. However, we keep the scale-dependent bias with $f_{\rm NL}^{\rm gal}$ for galaxies.

\subsection{Maximum SNR due to ISW}

Since the cross-correlation between CGWB anisotropies and LSS is dominated by the ISW effect, in order to estimate the maximum SNR, we will start by assuming that galaxies follow exactly the trace of the gravitational potentials \cite{Giannantonio:2008zi}.
For our case, a perfect trace implies that $C_{\ell}^{\text{CGW-gal}} \sim C_{\ell}^{\text{ISW}}$.
Under this approximation, one can estimate the maximum SNR as,
\begin{equation}
    \mathrm{SNR}^{2} = \sum_{\ell = \ell_{\rm min}}^{\ell_{\rm max}}(2\ell + 1) \frac{C_{\ell}^\text{ISW}}{C_{\ell}^\text{CGW}}.
\end{equation}
where $C_{\ell}^\text{ISW}$ and $C_{\ell}^\text{CGW}$ are given by Eqs. (\ref{eq:Cl_gwgw}) and (\ref{eq:Cl_isw-isw}), respectively.

In the last column of Table \ref{tab:snr}, we show the best possible SNR under the previous assumptions for some preferred values of $\locfnl$. 
The SNR values show that for this perfect scenario, we could potentially have a detection of the cross-correlation between CGWB and LSS for some selected values of $f_{\rm NL}$. To obtain more realistic estimates of detection we need to consider the actual cross-correlation of the GW anisotropies with galaxies.

\begin{table}[t]
\caption{\label{tab:snr}%
SNR values for the cross-correlation for different $\locfnl$ values.
Notice that in the last row we are making the distinction that we are using $f_{\rm NL}^{\rm gal}$ to differentiate it from the fixed $f_{\rm NL}^{\rm GW}=0$ that defines general CGWB anisotropies}
\begin{ruledtabular}
\begin{tabular}{l c c c}
\textrm{CGWB model} & \textrm{ET+CE} & \textrm{CV only} & \textrm{Max-ISW} \\
\colrule
\noalign{\vskip 2pt}
SIGW $f_{\rm NL} = -0.10$ & 3.74 & 16.82 & 24.77 \\
SIGW $f_{\rm NL} = 0.40$  & \textbf{1.00} & 2.59 & 4.36 \\
SIGW $f_{\rm NL} = 1.25$  & 0.47 & \textbf{1.00} & 1.71 \\
SIGW $f_{\rm NL} = 10$  & 0.10 & 0.14 & 0.23 \\
\noalign{\vskip 2pt}
\hline
\noalign{\vskip 2pt}
General CGWB $f_{\rm NL}^{\rm gal} = -0.1$ & 2.70 & 9.00 & 14.0 \\
General CGWB $f_{\rm NL}^{\rm gal} = 10$ & 2.97 & 9.11 & 14.0
\end{tabular}
\end{ruledtabular}
\end{table}
\subsection{SNR for CGWB and LSS cross-correlation}

When large-scale structures do not exactly trace the potential, we need to extend our analysis.
For a more realistic, but still simplified scenario, we adopt two simplifying assumptions.
First, we assume that for all our analysis the measurements are performed over the full sky, which implies $f_{\rm sky}=1$. Under full-sky consideration, for simplicity, we also neglect the effect of the integral constraint on large scales \cite{Riquelme:2022amo, Terasawa:2025qjl}.
Second, we assume that galaxy clustering is limited by cosmic variance alone, since shot-noise is negligible on the large scales for high-density surveys like LSST.

The previous considerations allow us to ignore correlations between different $\ell$ and focus only in the full variance, defined as,
\begin{align}\label{eq:variance}
    & \mathrm{Var}\left(C_\ell^\text{CGW-gal}\right) = \frac{1}{(2\ell + 1)
    f_{\mathrm{sky}}}  \\
    & \qquad \times  \left[ C_\ell^\text{gal} \left( C_\ell^\text{CGW} + N_\ell^\text{GW} \right) + \left( C_\ell^\text{CGW-gal} \right)^2 \right].\nonumber
\end{align}
For the previous equation, we use $C_\ell^\text{gal}$ as defined in Eq.(\ref{eq:cl_galgal}), $C_\ell^\text{CGW-gal}$ defined in Eq.(\ref{eq:cl_ggw}) and $C_\ell^\text{CGW}$ defined as Eq.(\ref{angular-gw-contrast}).

To study the importance of the GW detector noise against CV, for all the analysis in this section, we compare the different results for two cases: using only CV and the full noise (ET+CE+CV). 
\begin{figure}[t]\label{fig:snr_fnl}
  \centering
    \includegraphics[width=\columnwidth]{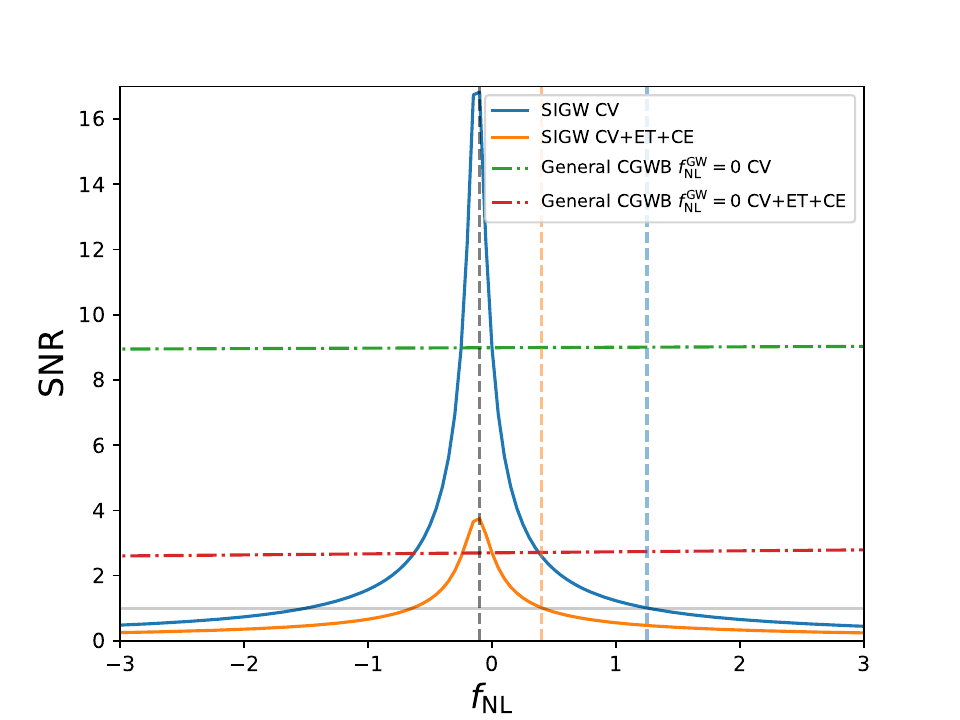}
  \caption{SNR for the cross-correlation as a function of $\locfnl$.In blue, we show the SNR of the same signal with the noise coming only from CV. In orange, is the SNR of the signal considering that the noise is combination of the GW detector noise and CV.  The vertical gray line shows the value of $f_{\rm NL}$ that has the better SNR. Vertical dashed blue line shows one of the critical points of our analysis, the value of $f_{\rm NL}$ that crosses $\text{SNR}\sim1$ for the case of CV only. On the other hand, the orange vertical dashed line shows which $f_{\rm NL}$ crosses $\text{SNR}\sim1$ for the CV+ET+CE. On the other hand, the horizontal lines correspond to a general CGWB anisotropy for which $f_{\rm NL}^{\rm gal}=0$.}
  \label{fig:snr_fnl}
\end{figure}
Using the previously defined variance, we can compute the SNR with the following definition:
\begin{equation}
    \mathrm{SNR}^2 = \sum_{\ell_\text{min}}^{\ell_\text{max}}\frac{\left(C_\ell^{\text{CGW-gal}}\right)^{2}}{\mathrm{Var}\left(C_\ell^{\text{CGW-gal}}\right)}.
\end{equation}
In Figure \ref{fig:snr_fnl}, we show how the SNR changes for different values of $\locfnl$.
Due to the arguments presented before, we compute the SNR for two configurations: CV only and CV plus GW noise.
The figure also highlights in vertical blue and black dashed lines two selected critical points, where $\text{SNR}\sim1$ for different error configurations. The vertical blue dashed line shows the value of $f_{\rm NL}$ at which the cross-correlation, in the presence of detector noise and CV, could be detected. On the other hand, the vertical black dashed line highlights the maximum value of $f_{\rm NL}$ that could have a detection over CV only.
The selected critical values will serve as fiducial for the forecast in the next section.

From the figure, we can see that the maximum of the SNR is peaked around $\locfnl=-0.1$ (vertical black dashed line) for both configurations.
For small negative values of $\locfnl$, the late-ISW becomes the main contribution in the autocorrelation of the CGWB anisotropies, as presented in the top panel of Figure \ref{fig:cls_comparison}, producing the peak that we see in the figure.
This suggests that accounting for even a slight non-Gaussianity enhances the SNR relative to the purely Gaussian case.
Interestingly, the location of this maximum SNR coincides with the latest analysis of {\scshape Planck} PR4 data \cite{Jung:2025nss}. However, given the uncertainty of the order $\sigma(f_{\rm NL})\sim 5$, this agreement should be treated with caution. 
In Table \ref{tab:snr}, we summarize the SNR values of the cross-correlation for the critical values of $\locfnl$.

Our results suggest that non-Gaussian SIGW offers prospects for a detectable $\text{LSS}\times\text{CGWB}$ correlation signal thanks to synergies between $f_{\rm NL}$ in both CGWB and LSS, highlighting the importance of including PNG in the modeling. This synergy also suggests that SIGW has better detectability prospects compared to other types of CGWB anisotropies where the initial condition does not correlate with LSS.
We also emphasize that in a cosmic-variance-only analysis, our conclusions will be independent of the pivot frequency and amplitude of the GW background, which only affect the recovery of the GW anisotropy signal from detector noise.

\subsection{Forecast of PNG}

After been established the possibility of detection of this cross-correlation and the importance of $\locfnl$.
One might wonder how this cross-correlation could improve the constraints of PNG when combined with galaxies.

The forecast is based on the Fisher matrix formalism, where, for a set of parameters $\theta_{i}$ and Gaussian likelihood, is defined as:

\begin{equation}
    F_{ij}=-\left\langle \frac{\partial^{2}\ln\mathcal{L}}{\partial \theta_{i}\partial \theta_{j}}\right\rangle = \sum_{\ell} \frac{1}{\mathrm{Var}_\ell\left(\mathbf{C}_{\ell}^{XY}\right)} \left( \frac{\partial \mathbf{C}_{\ell}^{XY}}{\partial f_{\mathrm{NL}}} \right)^2,
\end{equation}
where, as before, we ignore covariance between multipoles due to mask effects by setting $f_{\rm sky}=1$. 
Throughout this work we also ignore cross-correlation between redshift bins for the galaxy APS.
We consider that the forecast is only performed for the $\locfnl$ parameter, this should give us most of the constraining power by ignoring possible degeneracies between other parameters and $\locfnl$.
We leave the more realistic considerations of the analysis for future work, keeping in mind that realistic scenarios will worsen the results of this analysis.
From the Fisher matrix, the forecast error, $\sigma(f_{\rm NL})$, can be obtained as,
\begin{equation}
    \sigma(f_{\rm{NL}}) = \frac{1}{\sqrt{F_{f_{\rm NL}f_{\rm NL}}}}
\end{equation}

For the forecast, we use the critical values from the previous section as fiducial, $f_{\rm NL}^{\rm fid}=[-0.1, 0.4, 1.2]$.
We also perform the forecast for two variances, CV only and CV plus ET+CE noise.
For the fiducial galaxy clustering model only, we consider the LSST-like survey with the properties described in Section \ref{sec-III}. We construct a theory vector as follows,  
\begin{equation}
\mathbf{C}_\ell^{\rm gal} = \left(
C_\ell^\text{gal}(z_1),...,C_\ell^\text{gal}(z_i)\right),
\end{equation}
where $z_{i}$ corresponds to each redshift bin.
On the other hand, the fiducial cross-correlation is the one computed in Section \ref{sec-IV}, including the cross-correlation of LSS coming for each redshift bin, that is, the theory vector can be represented as,
\begin{equation}
\mathbf{C}_{\ell}^{\text{CGW-gal}} = \left(
C_\ell^\text{CGW-gal}(z_1),...,C_\ell^\text{CGW-gal}(z_i).
\right)
\end{equation}

In the spirit of a multi-tracer approach, we also perform a joint analysis of LSS and CGWB anisotropies to assess how their combination improves the constraints on $\locfnl$. We use fiducial full theory vector represented as,
\begin{equation}
\mathbf{C}_\ell = \left(
\mathbf{C}_\ell^{\rm gal} ,
\mathbf{C}_{\ell}^{\text{CGW-gal}}
\right).
\end{equation}
For each theory vector, we perform the forecast with each respective variance. For the combined constraints, from Figure \ref{fig:cls_comparison}, we show that the amplitude of the galaxy autocorrelation is $\sim9$ orders of magnitude larger that the cross-correlation, this implies that we can ignore the cross-covariance term and combine the Fisher information as follows,
\begin{equation}
    F^{\text{LSS}+\text{LSS$\times$CGWB}}_{f_{\rm NL}f_{\rm NL}} = F^{\text{LSS}}_{f_{\rm NL}f_{\rm NL}} + F^{\text{LSS$\times$CGWB}}_{f_{\rm NL}f_{\rm NL}}.
\end{equation}
In Table \ref{tab:forecast}, we show the forecast for the uncertainty in $f_{\rm NL}$ for different fiducial models. The $f_{\rm NL}$ values were selected based on the criteria that give the best SNR under certain considerations, as argued in the previous section.
\begin{table}[t]
\caption{\label{tab:forecast}%
Forecasted $\sigma(f_{\rm NL})$ values for the different correlations. Combined corresponds to LSS + LSS$\times$CGWB.}
\begin{ruledtabular}
\begin{tabular}{l c c c}
\textrm{Fiducial} & LSS & \( \text{LSS}\times\text{CGWB} \) & \( \text{Combined}\) \\
\colrule
\noalign{\vskip 2pt}
\textbf{$f_{\rm NL}=-0.1$}, CV only & 3.181 & \textbf{10.26} & \textbf{3.04} \\
\textbf{$f_{\rm NL}=-0.1$}, CV+ET+CE & 3.181 & \textbf{11.75} & \textbf{3.07} \\
$f_{\rm NL}=0.4$, CV only & 3.182 & 34.48 & 3.16 \\
$f_{\rm NL}=0.4$, CV+ET+CE & 3.182 & 43.05 & 3.173 \\
$f_{\rm NL}=1.2$, CV only & 3.186 & 43.76 & 3.178 \\
$f_{\rm NL}=1.2$, CV+ET+CE & 3.186 & 68.20 & 3.182 \\
\end{tabular}
\end{ruledtabular}
\end{table}

In this section, we present two main results. First, we show that for the critical values of $f_{\rm NL}$ ({\it i.e.,} the ones with better SNR in certain scenarios), we find that we can constrain $f_{\rm NL}$ by only using the cross-correlation of CGWB anisotropies and LSS.
More interestingly, we show that, by combining both the cross-correlation CGBW anisotropies and LSS along with the autocorrelation of galaxies, we are able to improve the constraints up to 4\%  for the case of $f_{\rm NL}^{\rm fid}=-0.1$ when considering only CV. The improvement for a more realistic case, ie. including the noise of the ET+CE detectors, is about a $3.5\%$. 

\section{Conclusions}\label{sec-VI}

In this article, we have shown that the large-scale structures of the universe (LSS) correlate with the cosmological gravitational wave background (CGWB). 
Concretely, we numerically computed the angular cross-correlation between the anisotropies of the CGWB and the galaxy density contrast for a LSST-like survey, where we showed that the correlation is dominated by the late-ISW effect of the tensor modes. We give an analytical estimate of the effect at large scales, where we found that the cross-correlation scales like $C_{\ell}^{\text{CGW-gal}} \sim C_{\ell}^{\text{ISW-gal}} \propto 1/\sqrt{(\ell +1)^{3}} + A\locfnl/\ell \sqrt{(\ell + 1)^{3}}$, with A constant. This differs from the expected signal for the cross-correlation of LSS and astrophysical GWB anisotropies, $C_\ell^{\text{AGW-gal}} \sim 1/(\ell+1/2)$, suggesting that the source of the SGWB anisotropies could be identified by comparing the behavior of the cross-correlation at large scales.

Furthermore, by considering a CGWB sourced by SIGWs and including a scale-dependent galaxy bias, we showed how primordial non-Gaussianity, by means of $\locfnl$, influences the overall amplitude of the cross-correlation. 
We performed a SNR analysis to study the detectability of these effects for future GW surveys such as the Einstein Telescope (ET) and the Cosmic Explorer (CE). We found that the SNR peaks at $\locfnl=-0.1$, suggesting that accounting for even a slight non-Gaussianity enhances the SNR relative to a general CGWB anisotropy.

In light of a possible detection, we performed a Fisher-matrix forecast for different values of $\locfnl$ in the model.
We found that the cross-correlation between SIGW anisotropies and LSS would be able to constrain in $\locfnl$, with a $\sigma(\locfnl)=10.26$ for the ideal cosmic-variance-only case, and $\sigma(\locfnl)=11.75$ when we include the noise of the ET and CE combined.
We found that, by performing a joint Fisher analysis, the combination of $\text{LSS}\times\text{CGWB}$ and galaxy-only LSS observables improves the constraints on $\locfnl$ by 4\% compared to using the galaxy angular power spectrum alone. This result highlights the potential of cross-correlating CGWB anisotropies with LSS as a complementary probe, offering a new avenue to enhance sensitivity to primordial non-Gaussianity in future cosmological surveys.

Further extensions from the observational cosmology side could include the study of the cross-correlation using other galaxy surveys such as Euclid \cite{Euclid:2024yrr}, DESI \cite{DESI:2019jxc}, to investigate the impact of the gravitational ISW effect in the context of spectroscopic surveys. Additionally, one could also study the effects of a galaxy survey footprint mask. For galaxy surveys, the access to large-scale modes is usually limited by survey mask effects; we consider these to be worth pursuing in the future since they affect our scales of interest.
On the other hand, the detection of GW anisotropies is limited by the properties of the GW detectors, and future experiments such as the Big Bang Observer \cite{Corbin:2005ny} and DECIGO \cite{Kawamura:2020pcg} could improve their detectability. This suggests that further research could include an intensive analysis of how the interplay with different detectors improves the detection prospects of the CGWB-LSS cross-correlation. 
Moreover, as argued before, both AGWB and CGWB anisotropies could have a different behavior at large scales, so their cross-correlation with LSS could be a way to tell them apart. This could open a new avenue of research involving the joint analysis of AGWB and CGWB anisotropies and their cosmological implications.

Previous works have analyzed the correlation between the AGWB with LSS, we trust that a complete analysis of the cross-correlation of the full SGWB with LSS, should include the result presented in this work. The effect of such inclusion would modify the status of the SGWB as an LSS tracer. 
\\

{\it Data availability:} The scripts to replicate the results of this work, along with the modifications of $\texttt{GW\_CLASS}$, will be shared upon publication in \url{https://github.com/walter-riquelme}.

\begin{acknowledgments}
We would like to thank Rogerio Rosenfeld for comments in the draft.
RB acknowledges support from the Fondecyt Postdoctorado Project No. 3240730 (ANID).
WR acknowledges support from the S\~ao Paulo Research Foundation (FAPESP), Brasil. Process Number 2023/07640-7.
\end{acknowledgments}

\bibliographystyle{JHEP}
\bibliography{bib}

\end{document}